\documentclass[12pt,final]{ias2}
\usepackage{epsfig,graphicx,float,wrapfig,color,amsmath}
\usepackage{subfigure}

\title{Effect of pairwise Dipole - Dipole interaction among three atoms Systems }

\author[]{P Anantha Lakshmi , Ashoka Vudayagiri\footnote{corresponding author,avsp@uohyd.ernet.in} {}  and Shaik Ahmed \\School of Physics, University of Hyderabad\\ Hyderabad - 500046, India}
\date{}
\begin{document}

\maketitle

\begin{abstract}
We present analysis of a system of three two-level atoms interacting with each other through the dipole-dipole interaction. The interaction manifests between excited state of one of the atoms and the ground state of its nearest neighbour. Steady state populations of the density matrix elements  are presented and are compared with a situation when only two atoms are present. It can be noticed that the third atom modifies the behaviour of the three atoms. Two configurations are analysed, one in which the three atoms are in a line, with no interaction between atoms at the end points and the other in which the atoms form a closed loop with each atom interacting with both  its neighbours. 
\end{abstract}

\begin{figure}[!h]
\centerline{\includegraphics[scale=0.2]{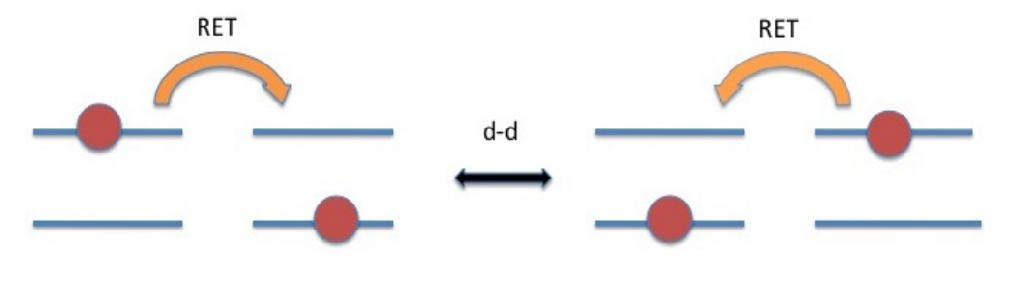}}
\caption{Resonant, non-radiative energy transfer between an excited state of one  atom and the ground state of the other}
\label{dipole-dipole}
\end{figure}

Whenever an atom in its excited state comes very close (within a  wavelength $\lambda$) to another atom in its ground state, energy is transferred to the second atom non-radiatively.  The second atom becomes the excited atom and the first one goes to ground state.  The phenomenon reverses after a while with the second atom transferring energy to the first one.  This kind of energy exchange  leads to an "interaction" between them.  The nature of this interaction is equivalent to the one derived from the dipole term of the classical electromagnetic interaction, and hence termed 'dipole-dipole' interaction  (see figure \ref{dipole-dipole}) \cite{cohen-tannoudji}. 

This simple coupling gives rise to a variety of phenomena, which have generated considerable interest in the past.   These include dipole blockade  \cite{comparat_blockade}, enhanced two-atom excitation  \cite{varada_agarwal}, or application of such situations for quantum computation \cite{quant_computation}  among others.  Further, when there are more than two atoms, wherein some of them are  excited while the others  are in their ground states, the dipole-dipole interaction acts between nearest ground-excited atom pairs. Many of the co-operative phenomena that are exhibited by a collection of atoms are manifestation of a coherent sum of these interactions. This provides motivation to extend these interactions to many atom systems.

The excitement to study dynamics of dipole-dipole interaction stems from the fact that this phenomenon can be exploited for several applications. For instance, Deutsch and coworkers, in the context of quantum information, have exploited the motion of a particle in double well to trap alkali aoms in a 1D lattice \cite{deutsch}.  They have employed a photon mediated collision resulting in an effective dipole-dipole interaction, which provided a handle to control an ensemble of uncoupled atomic systems.  They demonstrated that, controlling the strength of the optical field provides a direct control on the strength of the dipole-dipole interaction.   This is  perhaps one of the many ways of realizing the schemes that are studied in the present paper.

As a  first step, we extend the study to a set of three atoms. Three atoms in which any two interact with each other has also been of much interest. As a three-body problem, this leads to various effects such as  formation of Effimov states \cite{efimov} and/or correlation between those atoms that are not directly interacting with each other - mediated by the intermediate atom are some of the few phenomena that make the study of these  systems worthwhile. Some  of the studies are undertaken in the classical domain, using the classical dipole interaction picture \cite{martin}. Other studies exist  in the domain of van der Waals interaction instead of the above mentioned  dipole-dipole interaction\cite{axilrod,vanderwaal}.   Some of the recent works involve study of three atoms in a cavity \cite{cavity} and study of three fermionic atoms which form bipartite cooper pairs \cite{kensuke}. Results of `collisional' interaction between three atoms are also experimentally studied \cite{three_atom_collisions}. Yet, a simple study of dynamics that cover the entire range has not yet been undertaken.   In this  paper, we present the initial studies in this endeavour. The effects described here are generic and is valid for any atomic species, although they are more feasible in cold atoms in optical lattices. 

In the first section of this  communication, for the purposes of comparison, we present results for the case of  a system of  two atoms interacting with a common laser field.   When one of the two atoms gets excited, due to this field, a dipole-dipole interaction ensues between the two atoms. This  interaction is represented by a coupling factor `g'. Though the value of $g$ would critically depend upon the distance between the two atoms, we take it to be a constant - under the assumption that the inter - atomic distance is held constant.   Steady state solutions for the Liouville equation in the density matrix formalism 
are obtained for different values of field strength - represented by the corresponding Rabi frequency $\alpha$ and the dipole-dipole coupling factor $g$. Since the detailed results for the two atom case are published elsewhere\cite{pal_ahmed_ashok} , only highlights are recounted here.

In the next  section, three atoms, all of them interacting with the same field, are considered.  Inclusion of a third atom opens up multiple ways of arranging them.  But we consider two of the simplest and important configurations  - a linear chain where all the three atoms are on a single line and a closed  configuration, where all three are  on the vertices of an equilateral triangle(fig \ref{loopandline}). Any other arrangement would be a simple variation of these two.  In the linear (open loop) arrangement, the interaction between two farthest atoms can be neglected whereas in the closed loop arrangement,  all three atoms interact with each other with equal strength.  Because the atom-atom coupling manifests in different ways in both these configurations, the results for the two configurations are different as discussed in section 3.

\section{Two two-level atoms}
To formulate the problem in proper perspective, we at first present results of two atoms, which are interacting with each other via dipole-dipole interaction as well as an external electromagnetic field. The atomic response to this field is significantly altered by the dipole-dipole interaction, as shown below.

\begin{figure}[!h]
\includegraphics[scale=0.4]{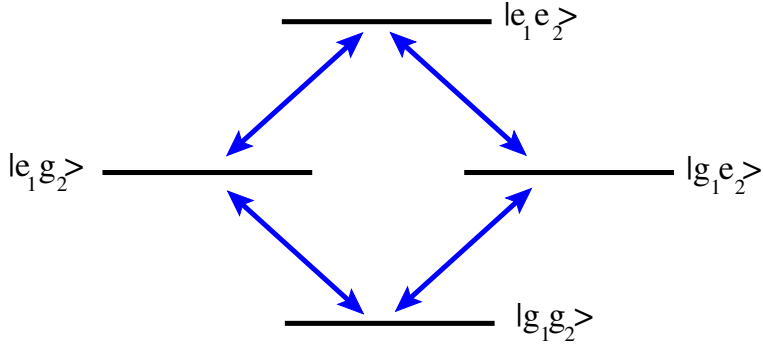}
\caption{The four level energy diagram}
\label{figure2}
\end{figure}

The energy levels of both atoms together can be represented in the combined Hilbert space of four energy levels as shown in figure \ref{figure2}. The state $|1\rangle$ represents state when both atoms are in ground state ($|g_1g_2\rangle$), state $|2\rangle$ and $|3\rangle$ represent states when either one of the atoms are excited while the other is still in ground state ($|g_1e_2 \rangle$ and $|e_1g_2 \rangle$) and the state $|4\rangle$ corresponds to the situation when both the atoms are in excited state ($|e_1e_2\rangle$).

The dipole-dipole interaction is modelled as a coupling factor `g' which brings about non-radiative  transitions between $|e_1g_2\rangle \leftrightarrow |g_1e_2\rangle$ or $|2\rangle \leftrightarrow |3\rangle$. The appropriate Hamiltonian for this system would be 

 \begin{equation}
 H=\begin{bmatrix}
\omega_{1} & -\alpha^{*} _{21} & -\alpha^{*} _{31} & 0 \\
-\alpha _{21} & \omega_{2} & -g & -\alpha^{*} _{42} \\
-\alpha_{31} & -g & \omega_{3} & -\alpha^{*} _{43} \\
0 & -\alpha _{42} & -\alpha _{43}& \omega_{4}\end{bmatrix} 
\end{equation}

which will be used to solve the Liouville equation 
\begin{equation}
i\hbar\frac{\partial\rho}{\partial t}=[H,\rho] + \mathcal{L} \rho
\label{Liouville}
\end{equation}

 The first term on the right hand side  represents the interaction with the radiation field and the dipole-dipole  interaction whereas the second term (The Liouvillean) accounts for different decay mechanisms. 

The resulting 16 equations, are reduced to 15 by using the completeness  condition $\rho_{11}+\rho_{22}+\rho_{33}+\rho_{44}=1 $, and rewritten in the form

\begin{equation}
\frac{\partial\Psi}{\partial t}=M\Psi+\Phi
\label{dynamics}
\end{equation} 
where $M$ is a $ 15 \times 15$  coefficient matrix  and   $\Psi$ and $\Phi$ are column vectors  each of length 15 which are defined in the following.

\begin{align} 
 &\Psi  = [ \rho_{11}~  \rho_{12}~  \rho_{13}~  \rho_{14}~  \rho_{21}~  \rho_{22}~  \rho_{23}~  \rho_{24}~  \rho_{31}~  \rho_{32}~  \rho_{33}~  \rho_{34}~  \rho_{41}~  \rho_{42}~  \rho_{43}]^T \\
 &\Phi   = [ 0~ 0~ 0~ 0~ 0~  2\gamma_{42}~ 0~ i \alpha^{*}_{42}~ 0~ 0~ 2\gamma_{43}~ i\alpha^{*}_{43}~ 0~ -i  \alpha_{42}~ -i\alpha_{43}]^T .
 \end{align}

The steady state solution of the density matrix elements may be  obtained by numerically solving for
\begin{equation}
\Psi_{ss} = \Psi(t\to\infty)=-M^{-1}\Phi
\label{steady_state}
\end{equation}

The detailed results of this system is presented elsewhere \cite{pal_ahmed_ashok}, but will be recalled here for the sake of completeness.

\begin{figure}[!h]
\subfigure[]{
\includegraphics[width=6.5 cm,height=5 cm]{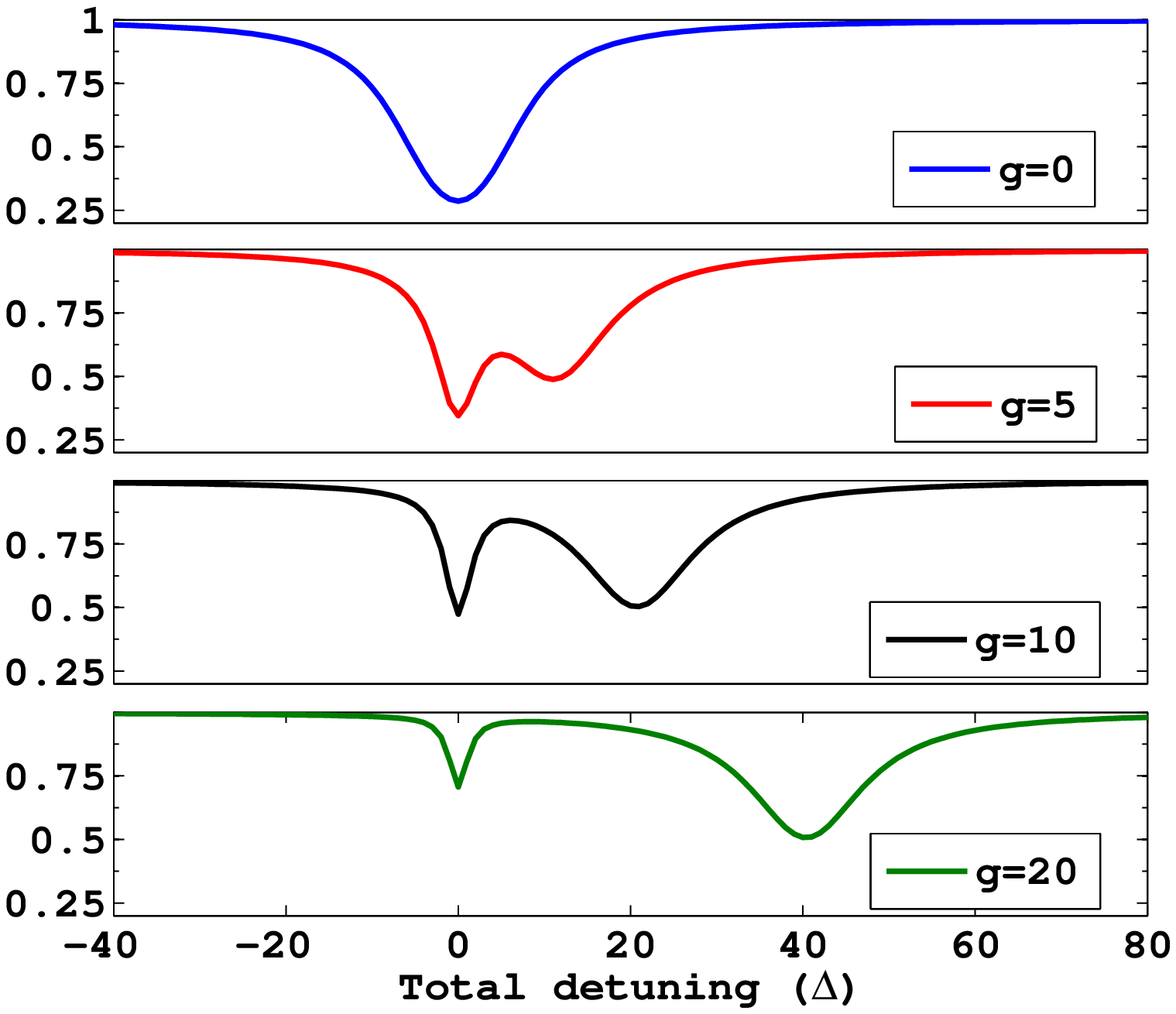}
\label{fig:subfig1}
}
\subfigure[]{
\includegraphics[width=6.5 cm,height=5 cm]{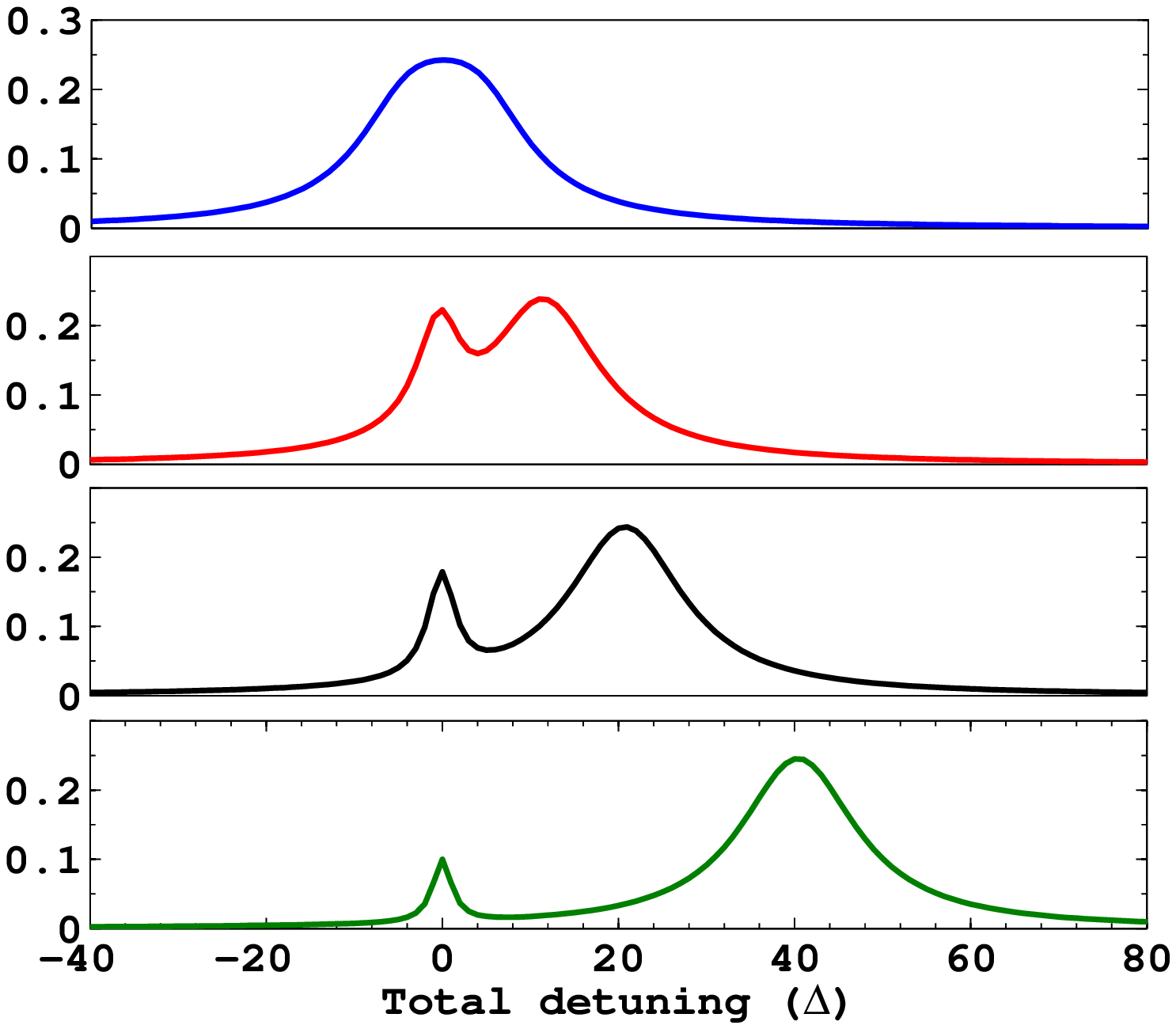}
\label{fig:subfig2}
}
\subfigure[]{
\includegraphics[width=6.5 cm,height=5 cm]{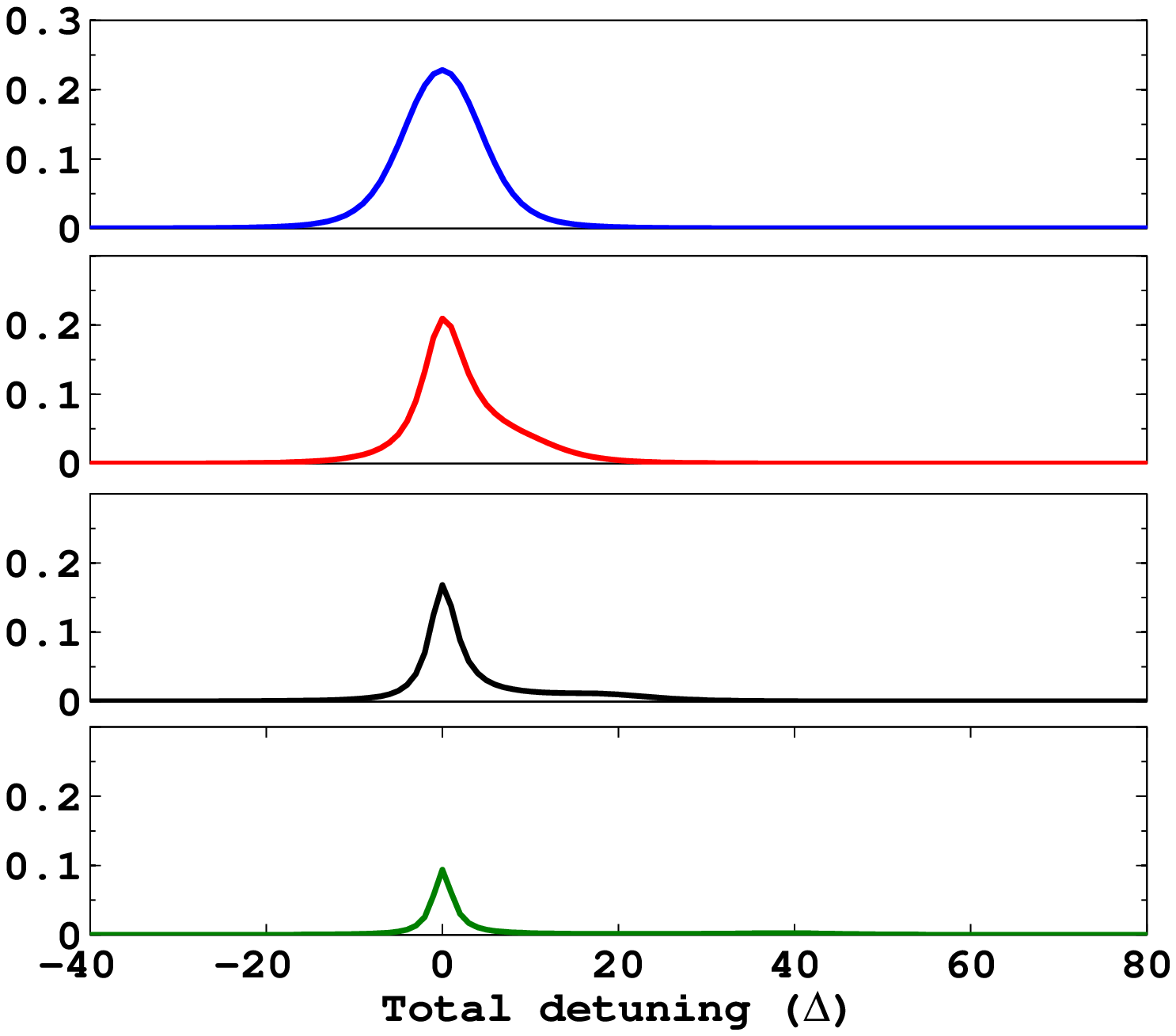}
\label{fig:subfig3}
}
\subfigure[]{
\includegraphics[width=6.5 cm,height=5 cm]{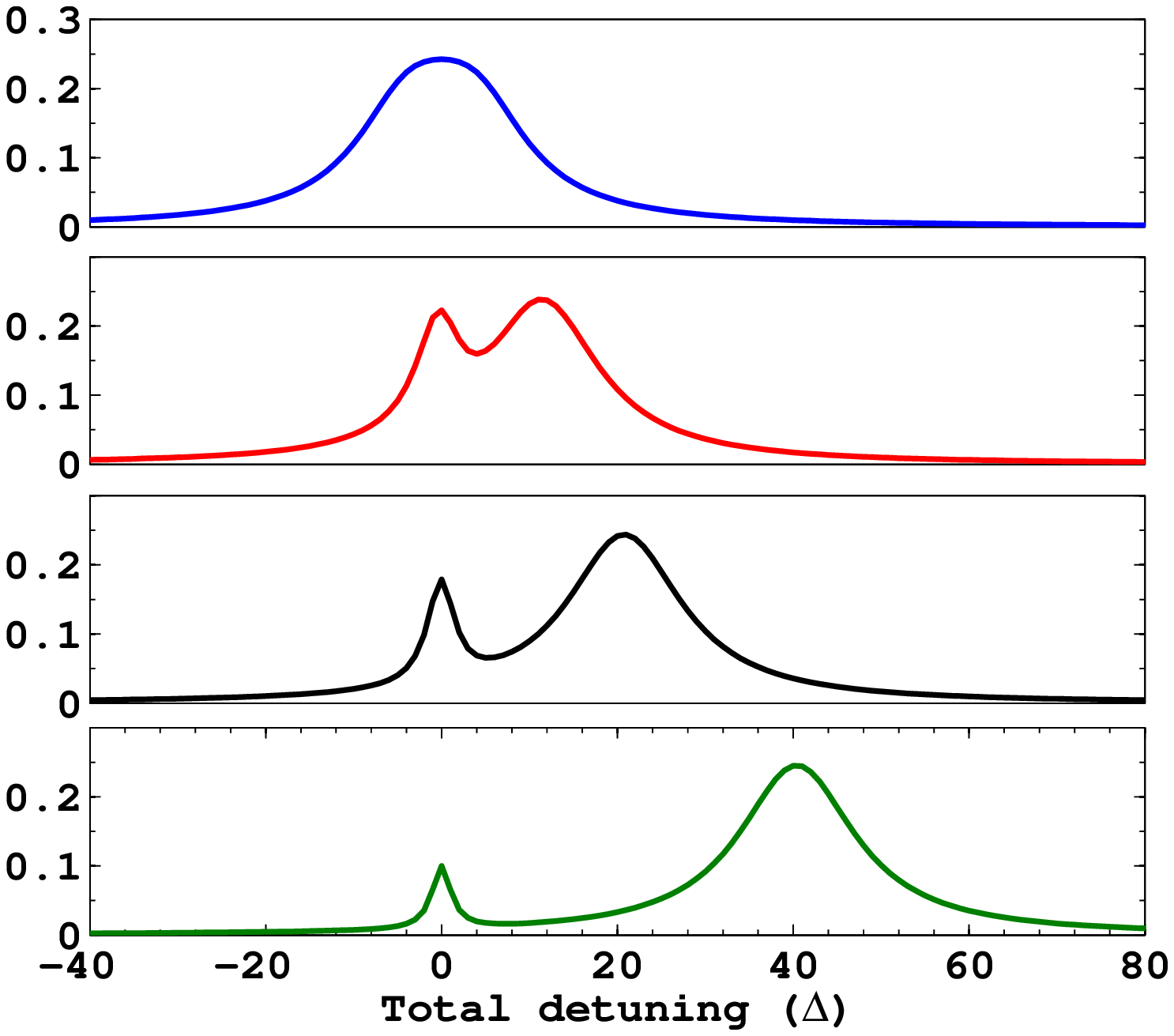}
\label{fig:subfig4}
}

\caption{Populations of levels $|1\rangle$, $|2\rangle$, $|3\rangle$ and $|4\rangle$ respectively in clockwise direction. For each subfigure, the coupling factors "g" are g = 0, 5, 10 and 20 top to bottom}
\end{figure}

Figures (\ref{fig:subfig1} - \ref{fig:subfig4}) show populations of the levels $|1\rangle$, $|2\rangle$, $|3\rangle$ and $|4\rangle$ respectively in clockwise direction. They are therefore populations of the coupled two - atom levels  $|gg\rangle$, $|ge\rangle$, $|eg\rangle$ and $|ee\rangle$. 

It can be seen that at $g=0$, the two atoms are independent of each other and hence the steady state population of both ground and excited states of both the atoms have a probability of $0.5$. Therefore, the probabilities of each level in the combined Hilbert space, will be product of these probabilities, i.e., $0.25$. As the coupling strength between the atoms increases, the energies of levels $|2\rangle$ and $|3\rangle$ undergo a shift and hence their populations show a sideband at the appropriate detuning $\Delta$. However this sideband is absent in the population of level $|4\rangle$, indicating  ``Dipole Blockade" wherein, once one of the atoms is excited while the other is still in ground state (either $|2\rangle$ or $|3\rangle$), it prevents the other atom getting excited. This can also be seen as a result of destructive interference between the pathways $|1\rangle \rightarrow |2\rangle \rightarrow |4\rangle$ and  $|1\rangle \rightarrow |3\rangle \rightarrow |4\rangle$. Nevertheless, a condition of both atoms simultaneously getting excited (bypassing levels $|2\rangle$ and $|3\rangle$) exists as indicated by the small peak of population of $|4\rangle$. The height of this peak reduces as a function of $g$, indicating a strong dependency on the coupling factor. 

When more than two atoms are involved, each atom interacts with all the others,  further complicating the system. However, one can make a simple extension, assuming that the atoms are located on equidistant lattice sites and only the nearest neighbour interaction is dominant.   The interaction with the other neighbouring atoms can be  neglected.  It can be understood that the effect of 'g' between any two neighbouring atoms is  same as shown earlier.   However the combined result  of the three - atom system is definitely different,  as shown in the next section. 

\section{Three atoms}
\begin{figure}[!h]
\subfigure[]{
\includegraphics[scale=0.15]{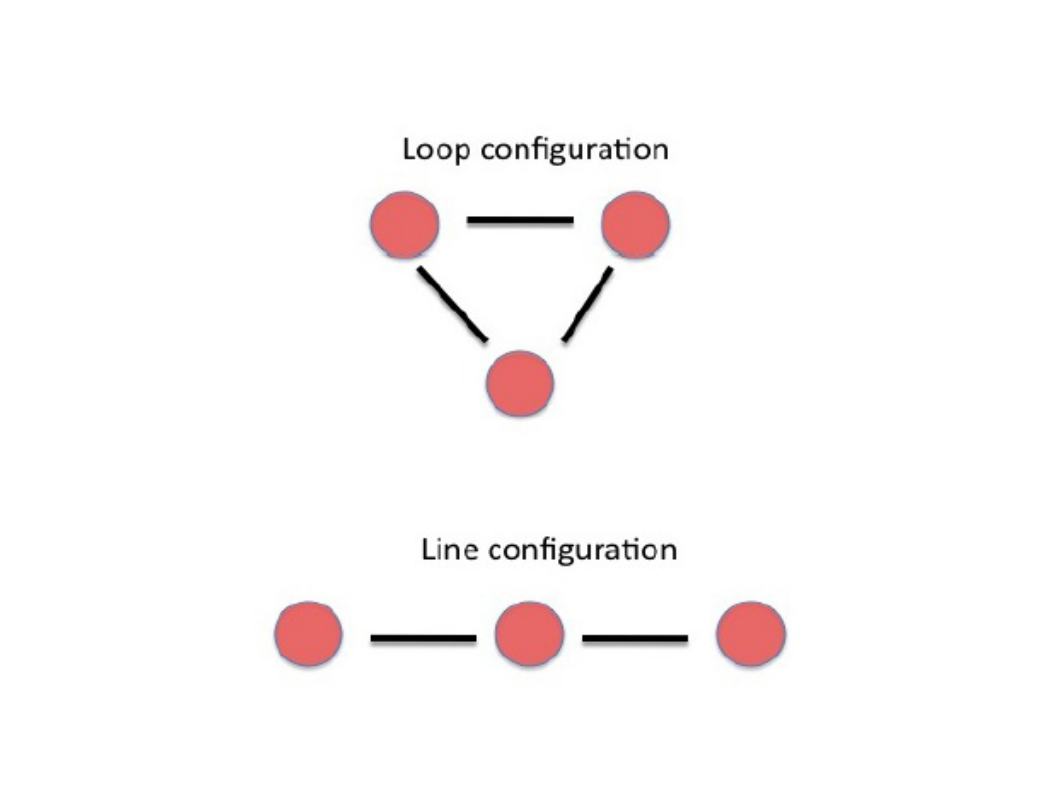}
\label{loopandline}
}
\subfigure[]{
\includegraphics[scale=0.2]{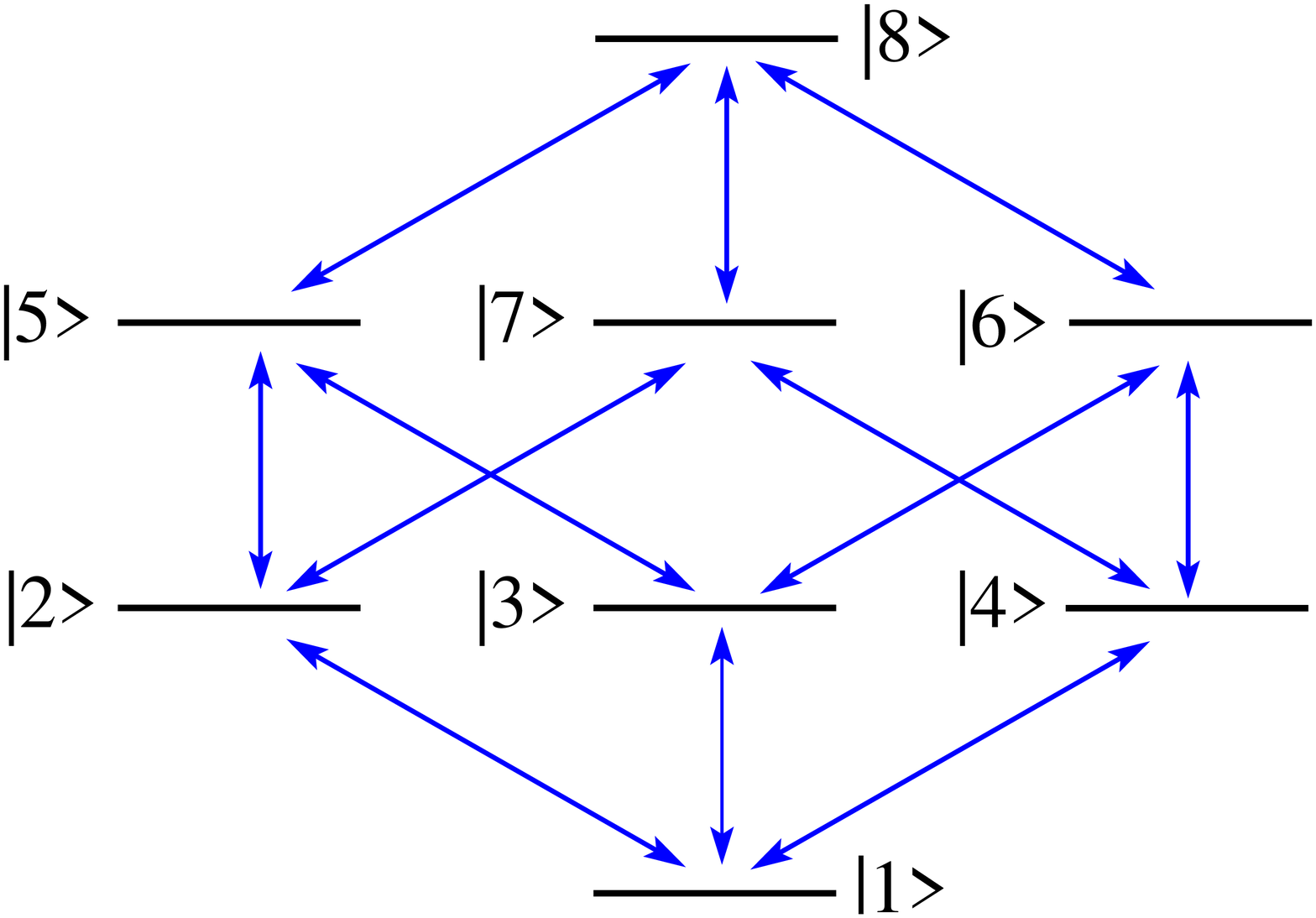}
\label{eight_levels}
}
\caption{(a) The closed loop (above) and line (open loop) arrangement (below) (b) The eight energy levels in the combined space. See text for definition of levels $|1\rangle$ - $|8\rangle$.}
\end{figure}

The above arguments will now be extended to three atoms, with bipartite d-d interactions. There are two configurations in which the three atoms may be arranged. In the  first  of these,  the atoms are arranged in a linear array, in which case d-d interactions exist between atoms 1 and 2, between atoms 2 and 3, but no interaction exists between atoms 3 and 1. The second arrangement is in the form of a  closed loop, where each of the three atoms interacts with its neighbours. The behaviour of the open-loop (linear) case is different from that of the closed loop. The two configurations are shown in figure \ref{loopandline}.

The energy levels in the combined space will form a system of eight energy levels, as shown in figure \ref{eight_levels}. Energy level labelled $|1\rangle$ corresponds to  $|ggg\rangle$ where all three  atoms are in ground state. The three levels wherein any one of the three atoms are excited are $|egg\rangle,~ |geg\rangle$ and $|gge\rangle$, which are all degenerate and are denoted respectively as $|2\rangle, |3\rangle$ and $|4\rangle$. The three  levels with two atoms in excited state and one in ground state are $|gee\rangle,~ |ege \rangle$ and $|eeg \rangle$ and are denoted $|5\rangle,~|6\rangle$ and $|7\rangle$. Finally the state with all three atoms excited is $|eee\rangle$ is denoted $|8\rangle$.  Figure \ref{eight_levels} also indicates the relevant laser couplings (blue arrows online) between the levels. This energy level scheme is same for both open loop and closed loop arrangements, except that for the open loop $g_{42}=0=g_{57}$, indicating that there is no dipole dipole interaction between atom 1 and atom 3.  

The Hamiltonian, including both laser coupling as well as the d-d coupling between different levels is given by
\begin{equation}
 H=\begin{bmatrix}
\omega_{1} & -\alpha^{*} _{21} & -\alpha^{*} _{31} & -\alpha^\ast_{41} & 0  & 0 & 0 & 0 & \\
-\alpha _{21} & \omega_{2} & -g_{23} & g_{24} & -\alpha^{*} _{52}  & 0  & -\alpha^\ast_{72} & 0  \\
-\alpha_{31} & -g_{23} & \omega_{3} & g_{34} & -\alpha^{*}_{53} & -\alpha^{*}_{63} & 0 & 0  \\
-\alpha_{41} & -g_{24} & g_{34} & \omega_{4} & 0 & -\alpha^{*}_{46} & -\alpha^{*}_{47} &  0  \\
 0 & -\alpha _{52} & -\alpha _{53}& 0 & \omega_{5}& g_{56} & g_{57} & \alpha^\ast_{85} \\
0 & 0  & -\alpha_{63} & -\alpha_{64} & g_{56} & \omega_6 & g_{67}  & -\alpha^\ast_{86} \\
0 & -\alpha_{72} & 0 & -\alpha_{74} & g_{57} & g_{67} & \omega_7  & -\alpha^\ast_{87}\\
0 & 0 & 0  & 0 & -\alpha_{85}  & -\alpha_{86}  & -\alpha_{87} & \omega_8 \\
\end{bmatrix} 
\label{eight_hamiltonian}
\end{equation}
in units of $\hbar$, with the usual notations.  For  the open-loop (single line) configuration,  the dipole coupling parameters  $g_{24}=0=g_{57}$ and   for the  closed loop configuration, all the  $g_{ij}$s  are nonzero. As a simplest case scenario, all the non zero  dipole-dipole coupling factors $g_{ij}$  both for the  open loop and  closed loop configurations are taken to be equal for computing  the dynamics of the system. The  Liouville equation for the density operator [\ref{Liouville}], governing the dynamics of the system, will give rise to  sixty four  coupled first order differential equations.  As is the usual practice, introducing the completeness condition $\sum_{i=1}^8 \rho_{ii}=1$ and eliminating one of the populations, say  $\rho_{88}$ in this study, results in a  set of sixty three coupled equations, which are solved in  steady state using same idea as outlined in the two-atom case. 

Intuitively, it can be noticed that the dipole-dipole interaction in case of closed loop configuration is identical among all the atom pairs.  Whereas, in the open loop configuration, the atom in the middle (atom 2) is interacting with two neighbours (1 and 3) whereas atoms 1 and 3 are interacting with only one neighbour(atom 2).  This lack of equivalence introduces difference in the behaviour of this system as compared to the closed loop system.   The results presented are clearly indicative of this behaviour.

 \section{Results and Discussion}
\begin{figure}
\centerline{\includegraphics[scale=0.6]{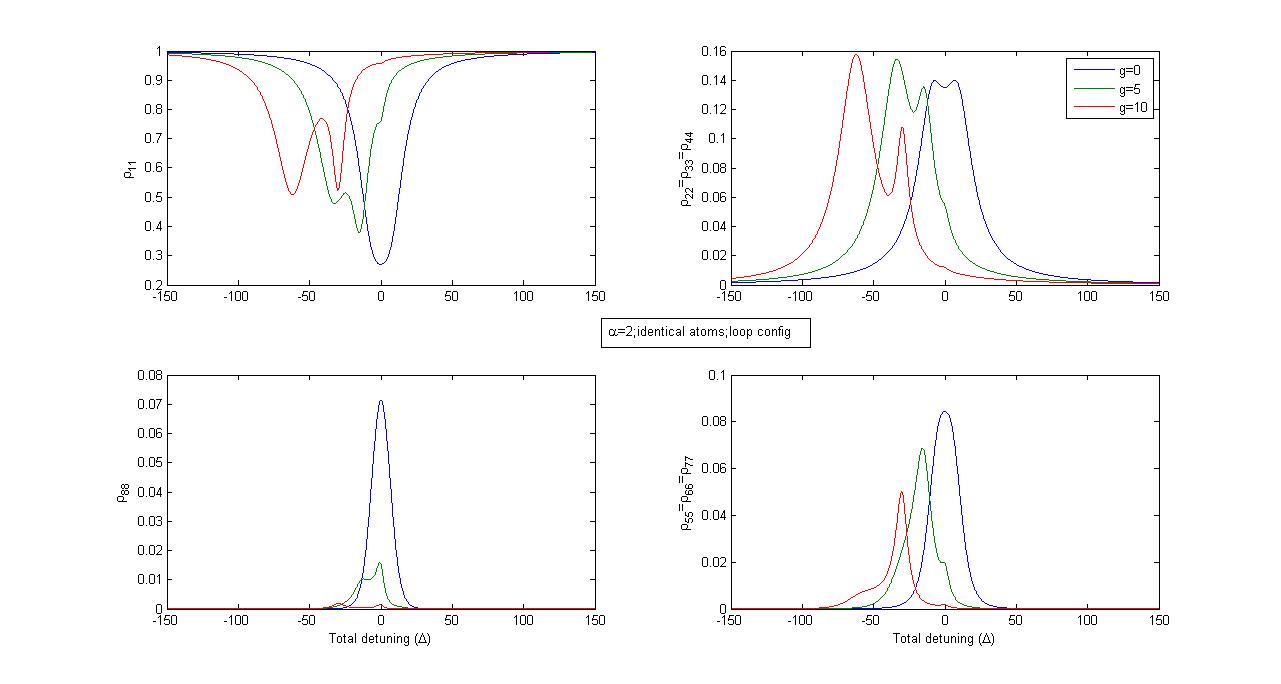}}
\caption{Populations for levels $|1\rangle$ to $|4\rangle$ for the loop configuration, for values of g=0 (blue curve), g=5 (green curve) and g=10 (red curve) (colors online). All curves for Rabi frequency $\alpha=5$}
\label{loop_populations1}
\end{figure}

The figures ( \ref{loop_populations1}) show plot of populations $\rho_{11}$ to  $\rho_{88}$, for the closed loop  configuration for different values of g. 
\subsection{Closed Loop configuration}

When $g=0$, $\rho_{11}$ shows a single dip at $\Delta=0$. However, the single excited atom case --  $\rho_{22}=\rho_{33}=\rho_{44}$ show two small peaks, instead of a single corresponding peak. The dip at the center can not be attributed to dipole blockade since at this point $g=0$. Instead, it has to be attributed to a loss of population to higher states, where more than one atom is excited. This is evident by the single peak structure seen in the  populations $\rho_{55}=\rho_{66}=\rho_{88}$. Unlike the two atom case, where all the populations were equal, it can be noticed that the system of three identical atoms shows different behaviour.  Individual probabilities for getting any one of the atoms in the excited state  (i.e., $\rho_{22}=\rho_{33}=\rho_{44}$) is equal to 0.14, which adds up to nearly a half of the  probability for the three one-atom-excited states. On the other hand, the case of any two of the atoms excited is much lower,  adding upto 0.24 for all the three states $\rho_{55}=\rho_{66}=\rho_{77}$ combined together. Probability for having all three atoms excited together is even smaller, which is about 0.08 approximately. 

Presence of $g$ affects the above situation in an interesting way. The dip for $\rho_{11}$ splits into two, with both the lobes shifted to red side of the resonance. This indicates that $g$ causes a mixing of the energy levels in such a way to create two superpositions, both shifted closer to each other. The two lobes also have asymmetric widths, with the extreme one becoming broader and stronger as $g$ increases. This behaviour is exactly mirrored by the two peaks for one-atom excited state $\rho_{22}=\rho_{33}=\rho_{44}$.

On the other hand, the two-excited-atoms  state does not exhibit a clear two-lobed structure. It shows a dominant single peak, which is matched at the resonance  to the narrower peak of the one-excited-atom case. For moderate $g$'s, a small bump can be seen at $\Delta=0$. This is most likely the population that has decayed from level $|8\rangle$, which has a peak at  $\Delta=0$. Level $|8\rangle$, which corresponds to all three atoms in excited state has only a sharp peak at the center, with no sidebands. The central peak too decreases drastically in height with increasing $g$. This can be interpreted as the presence of  dipole blockade in the case of three atoms, wherein the excited atoms prevent other atoms from getting excited. Presence of the  third atom indicates that the two-atom dipole blockade is not very effective, once  at least one of the atoms is excited.  Intuitively, one can then explain the two peaks of $\rho_{22},~\rho_{33}$ and $\rho_{44}$, and a single sideband of $\rho_{55},~\rho_{66}$ and $\rho_{77}$ as follows - the energy shift due to $g$ causes two resonances for atomic excitation - leading to any one of the three atoms to reach their respective excited states. Once excited, the atom is preventing  one of its neighbours from getting excited, which can be attributed  to absence of the broad peak in the populations.   However, the third atom is not affected drastically by this blockade and gets excited. In other words, the standard dipole blockade prevents only one of the two atoms from getting excited,  resulting in two atoms which can get excited.

\begin{figure}[!h]
\centerline{\includegraphics[scale=0.5]{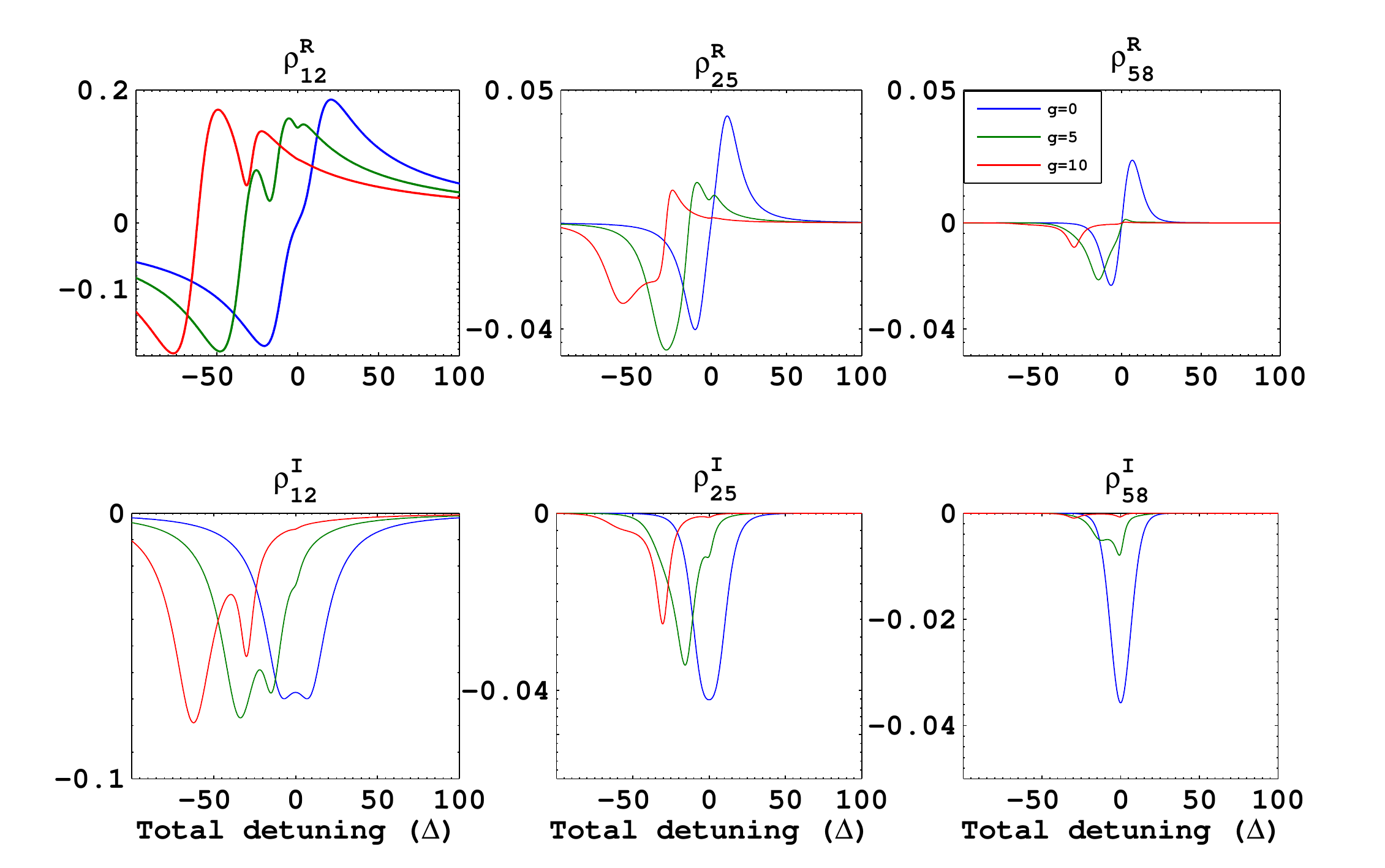}}
\caption{Real (row above) and Imaginary (row below) part of coherences $\rho_{12}$, $\rho_{25}$ and $\rho_{58}$ as labelled. For the loop configuration, $\rho_{12}=\rho_{13}=\rho_{14}$. Coherence $\rho_{25}=\rho_{27}=\rho_{35}=\rho_{36}=\rho_{46}=\rho_{47}$. And $\rho_{58}=\rho_{68}=\rho_{78}$.}
\label{first_coherence}
\end{figure}

Imaginary part of coherences, for case of dipole coupled transition,  indicate absorption of light. Figure \ref{first_coherence} shows these coherences for $\rho_{12}$, $\rho_{25}$ and $\rho_{58}$. The coherence $\rho_{12}$, which is also same as $\rho_{13}$ and $\rho_{14}$ for the loop configuration case shows absorption of one photon by one of the atoms to get excited. The populations $\rho_{22},~\rho_{33}$ and $\rho_{44}$ mirror this absorption profile exactly. Similarly, population of two-atom excited states $\rho_{55} -\rho_{77}$ and coherences $\rho_{25}$ (which is also equal to $\rho_{26},~\rho_{27}~\rho_{35}~\rho_{36}~\rho_{46}$ and $\rho_{47}$) mirror each other perfectly. Similarly, $\rho_{58}$ (=$\rho_{68}=\rho_{78}$) and the population term $\rho_{88}$ mirror each other. 

\begin{figure}[!h]
\centerline{\includegraphics[height=4cm,width=10cm]{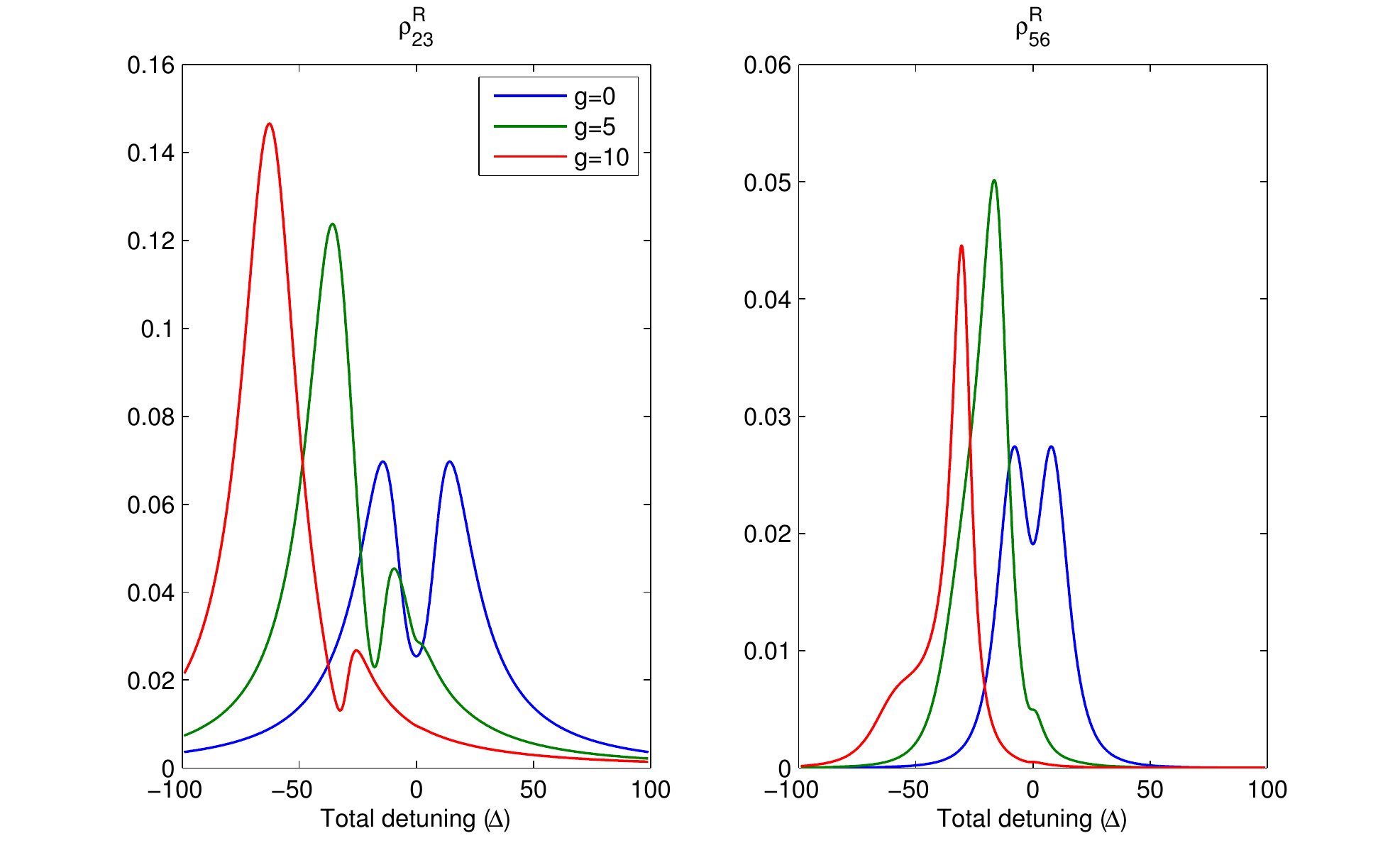}}
\caption{Real part of coherences :  $\rho_{23}=\rho_{23}=\rho_{24}$ and $\rho_{56}=\rho_{57}=\rho_{67}$ for loop configuration}
\label{second_coherence}
\end{figure}
\begin{figure}[!h]
\centerline{\includegraphics[height=4cm,width=10cm]{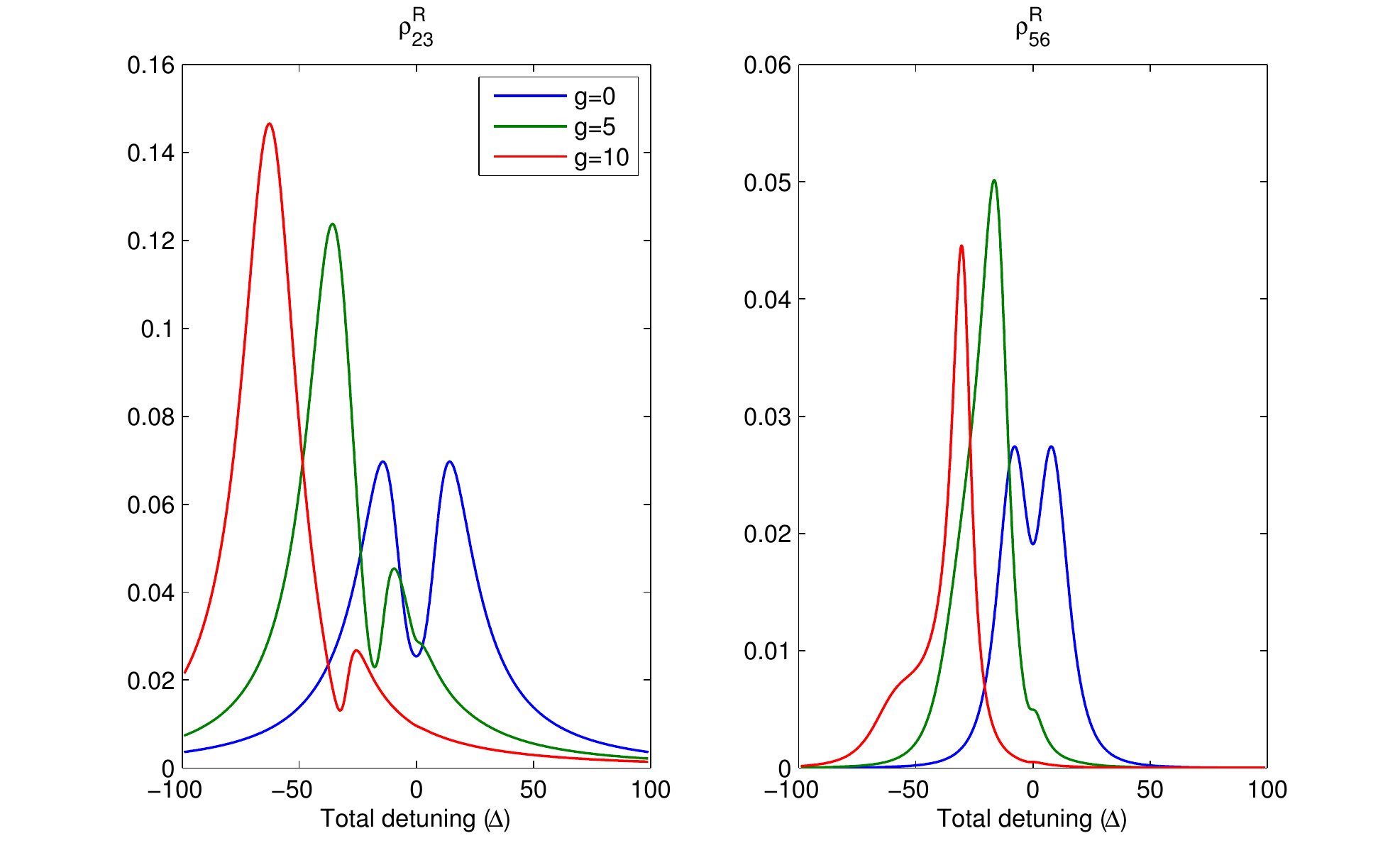}}
\caption{Real part of coherences : $\rho_{23}=\rho_{23}=\rho_{24}$ and $\rho_{56}=\rho_{57}=\rho_{67}$ for loop configuration}
\label{second_coherence}
\end{figure}

\subsection{Open Loop (line) Configuration}
Open loop configuration \ref{loopandline}, shows drastically different results. In this configuration, we label the state which corresponds to the middle , as  $|3\rangle$. This means that the dipole-dipole interaction couples states $|2\rangle$ to $|3\rangle$ and $|3\rangle$ and $|4\rangle$, but there is no coupling between $|2\rangle$ and $|4\rangle$. Similarly, there is no dipole-dipole interaction between states $|5\rangle$ and $|7\rangle$. 

\begin{figure}[!h]
\centerline{\includegraphics[scale=0.5]{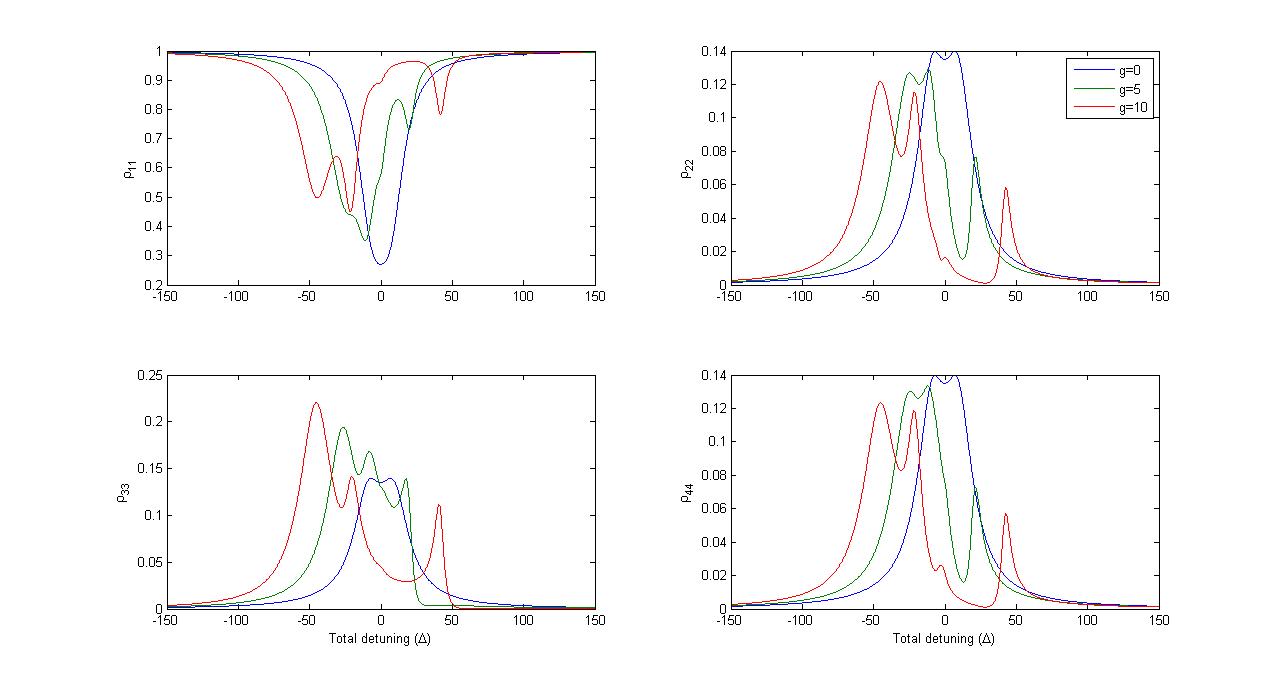}}
\caption{Populations for levels $|1\rangle$ to $|4\rangle$ for the line configuration }
\label{line_populations1}
\end{figure}

\begin{figure}[!h]
\centerline{\includegraphics[scale=0.5]{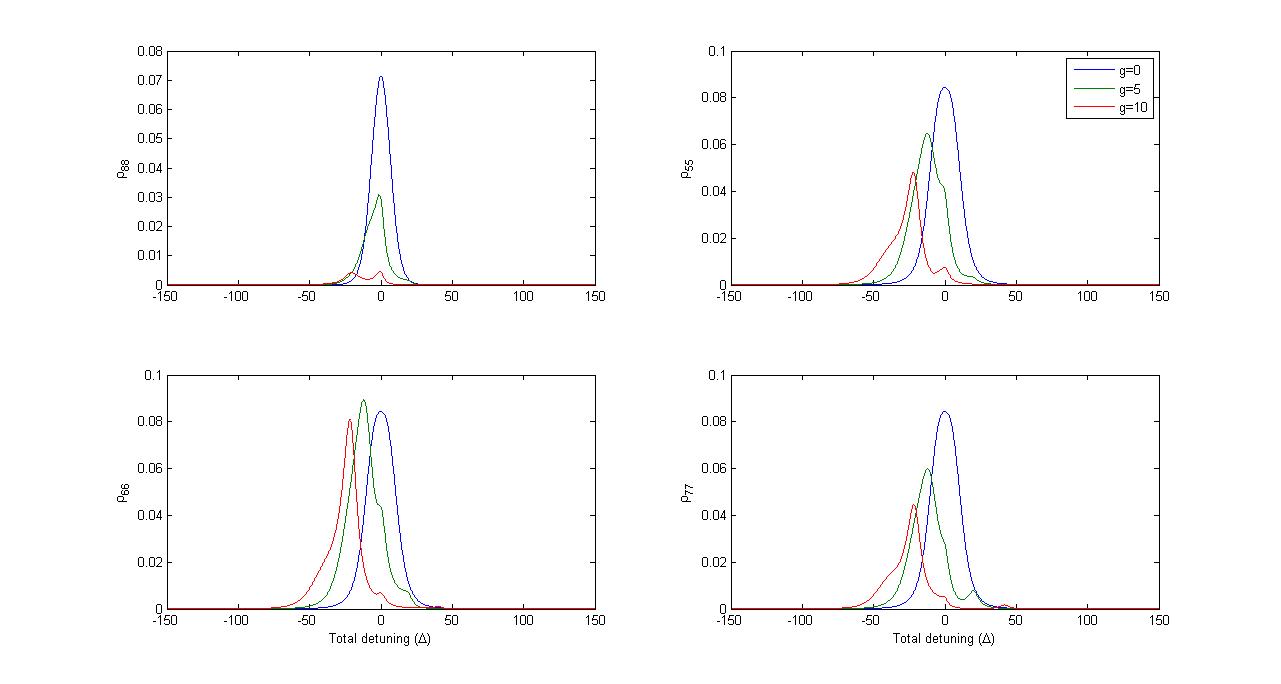}}
\caption{Populations for levels $|5\rangle$ to $|8\rangle$ for the line configuration }
\label{line_populations2}
\end{figure}

\begin{figure}[!h]
\centerline{\includegraphics[scale=0.5]{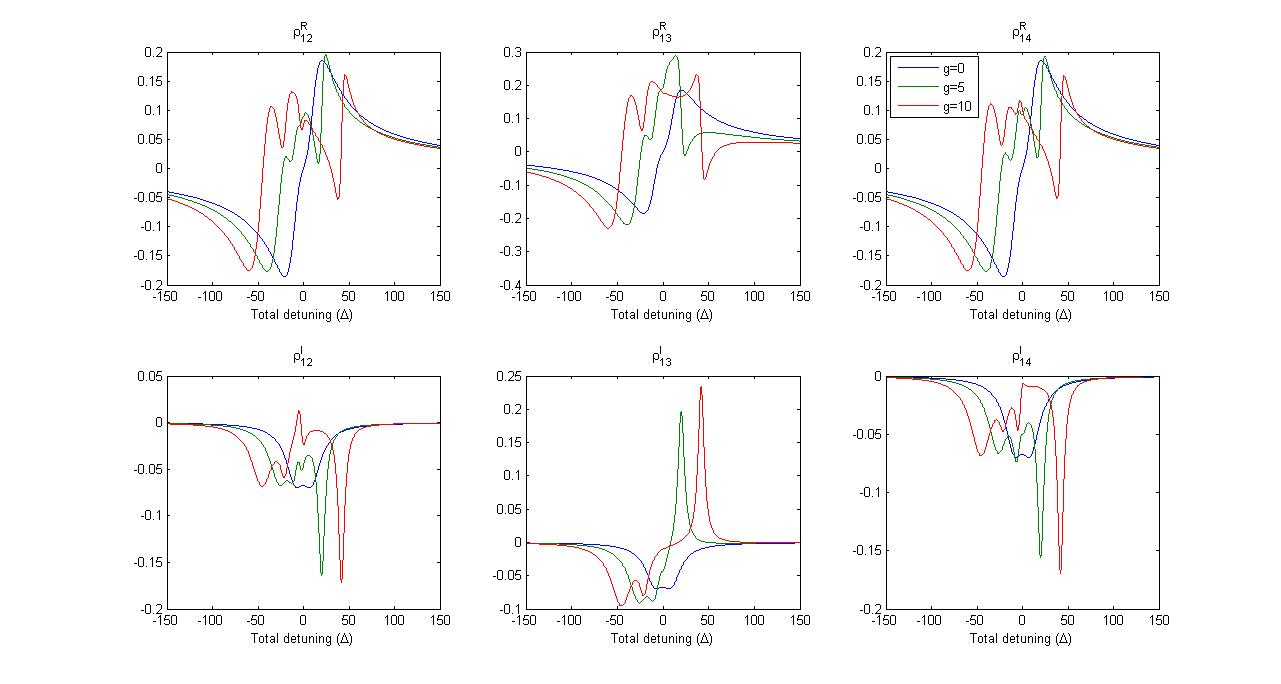}}
\caption{Coherences between one-atom excited states : $\rho_{12}$, $\rho_{13}$ and $\rho_{14}$  for the line configuration  }
\label{line_coherences1}
\end{figure}

\begin{figure}[!h]
\centerline{\includegraphics[scale=0.5]{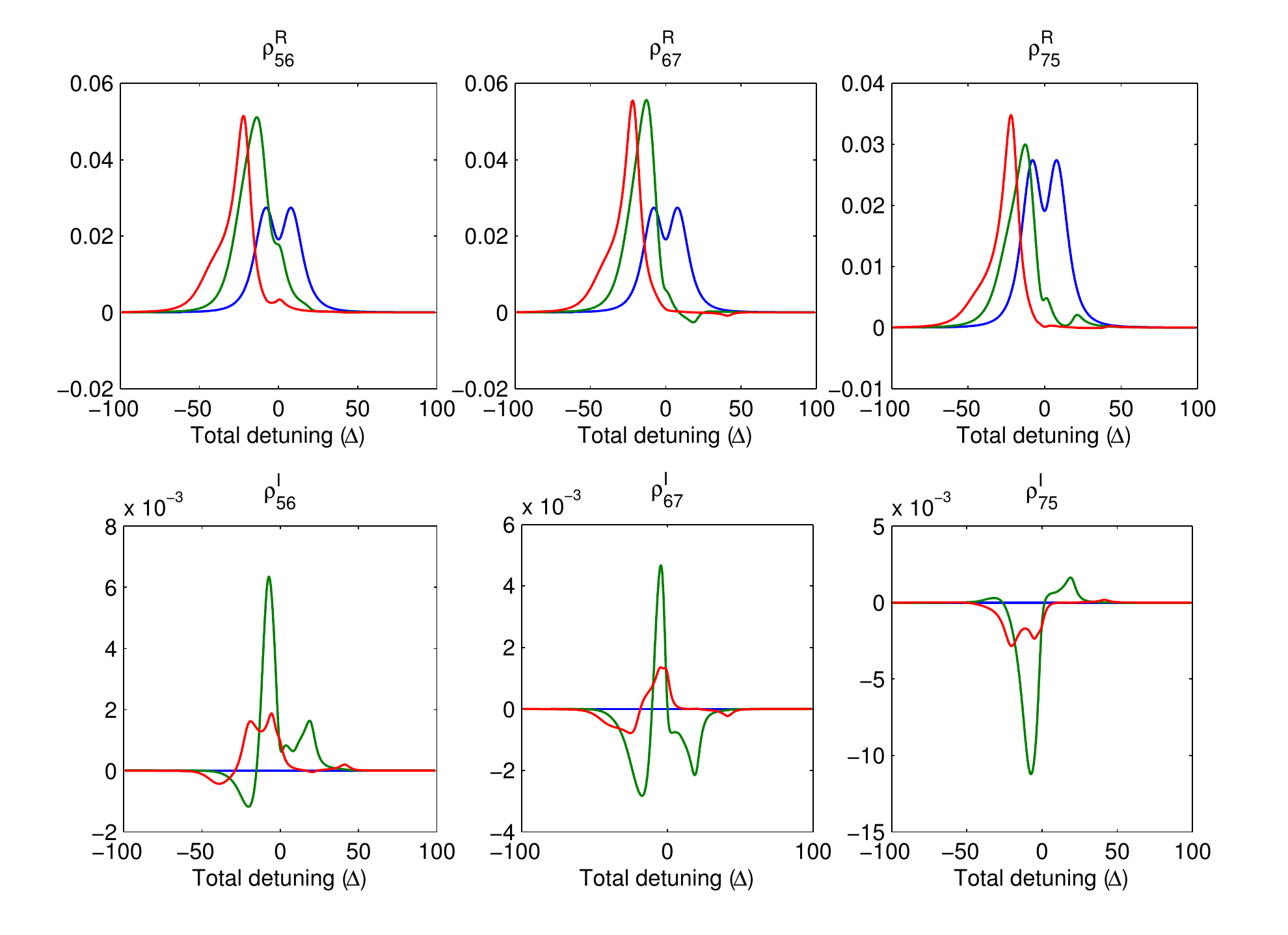}}
\caption{Coherences between two-atom excited states : $\rho_{56}$, $\rho_{67}$ and $\rho_{75}$  for the line configuration}
\label{line_coherences2}
\end{figure}
\vskip2.5cm

Obviously, one expects that the populations $\rho_{22}$ and $\rho_{44}$, which are both coupled to the state $|3\rangle$ through the dipole interaction, show identical behaviour while that of $\rho_{33}$ differs from these two.  For similar reasons one expects that $\rho_{55}$ and $\rho_{77}$ would be identical with each other, but different from that of $\rho_{66}$. The graphs shown in figure \ref{line_populations1} and \ref{line_populations2}  clearly indicate this expected behaviour.   There are three resonance peaks in absorption, as opposed to two that were present  in the loop configuration.  A small bump  at  the line centre ($\Delta=0$)  can be noticed in all the populations, indicating a three photon resonance.  The population $\rho_{88}$, on the other hand, starts with a single peak for $g=0$, which splits into two as $g$ is increased. The absorption is also strongly suppressed, indicating the presence of dipole blockade.

\section{Conclusion}
We have studied dynamics of three identical atoms interacting with each other via dipole-dipole interaction and also an external electromagnetic field. The dipole-dipole interaction is represented by a coupling factor `g' between an excited atom and an atom in the ground state. Considering only nearest neighbour interaction, we found that the interaction suppresses one of the neighbouring atoms from getting excited but not both.   We have studied two possible scenarios in which the three atoms can be present and the results for both these scenarios show different behaviour.   Attempt has been made to explain the differences in behaviour of the system under the two cases.  This study throws light on the general behaviour of the system but further insight can be obtained by refining the study.

\end{document}